\documentclass[amsmath,amssymb,prl,superscriptaddress,reprint]{revtex4-1}
\usepackage{graphicx} 
\usepackage{dcolumn} 
\usepackage{bm} 
\usepackage{hyperref} 
\usepackage{longtable} 
\usepackage[T1]{fontenc}
\usepackage{times}
\usepackage{xcolor}
\usepackage{wrapfig}
\usepackage{float}
\usepackage[inline]{enumitem}
\usepackage[addedmarkup=default]{changes}
\setaddedmarkup{\color{black}{#1}}
\graphicspath{{figures/}}

\newcommand{\Zcal}{\ensuremath{\mathcal{Z}}}

\begin{document}

\title{Relevance of Packing to Colloidal Self-Assembly}

\author{Rose \surname{Cersonsky}}
\thanks{These two authors contributed equally to this work.}
\affiliation{Macromolecular Science and Engineering Program, University of
 Michigan, Ann Arbor MI 48109, USA}
\author{Greg \surname{van Anders}}
\thanks{These two authors contributed equally to this work.}
\affiliation{Department of Chemical Engineering, University of Michigan, Ann
Arbor MI 48109, USA}
\author{Paul M.\ \surname{Dodd}}
\affiliation{Department of Chemical Engineering, University of Michigan, Ann
Arbor MI 48109, USA}
\author{Sharon C. \surname{Glotzer}}
\affiliation{Department of Macromolecular Science and Engineering, University of
 Michigan, Ann Arbor MI 48109, USA}
\affiliation{Department of Chemical Engineering, University of Michigan, Ann
Arbor MI 48109, USA}
\affiliation{Department of Materials Science and Engineering, University of
Michigan, Ann Arbor MI 48109, USA}
\affiliation{Biointerfaces Institute, University of
Michigan, Ann Arbor MI 48109, USA}

\date{\today}

\begin{abstract}
Since the 1920s, packing arguments have been used to rationalize crystal structures in systems ranging from atomic mixtures to colloidal crystals. 
Packing arguments have recently been applied to complex nanoparticle structures, where they often, but not always, work. 
We examine when, if ever, packing is a causal mechanism in hard particle approximations of colloidal crystals. We investigate three crystal structures comprised of their ideal packing shapes. We show that, contrary to expectations, the ordering mechanism cannot be packing, even when the thermodynamically self-assembled structure is the same as that of the densest packing. We also show that the best particle shapes for hard particle colloidal crystals in the infinite pressure limit are imperfect versions of the ideal packing shape. 
\end{abstract}
\maketitle

Why do atoms, molecules, or nanoparticles form the crystals they form? In 1929, Pauling proposed an answer to this question for atoms by demonstrating remarkable correlations between the sphere packing problem, the study of which dates back to Sanskrit writings in 499CE \cite{hales}, and the crystal structures of ionic solids \cite{paulingpacking}. The packing problem asks: given a set of hard, convex objects, such as spheres, what is the spatial
arrangement of those objects that most densely fills space? Pauling argued that crystal structures could be explained by packings of spheres of appropriate atomic radii.

Variants of the packing problem have yielded solutions relevant not only to the rationalization of crystal structures \cite{paulingpacking}, but also in optimal
information transmission \cite{Bischoff2002}, DNA in cell
nuclei \cite{Cremer2001,Marenduzzo2010}, blood clots \cite{polyhedrocytes}, plant
morphology \cite{BallTapestry}, and the stacking of oranges in the produce section \cite{halespackhist}. Packing rules were used by Frank and Kasper to rationalize
complex crystal structures in intermetallic alloys \cite{frankkasper1,frankkasper2}, and the molecular packing parameter, a popular geometric measure in surfactant self-assembly, is also based on packing principles \cite{israelachvilipf}.

More recently, Pauling's packing principles have been used to rationalize and predict colloidal crystals and nanoparticle superlattice structures by asserting packing as a causal mechanism. For example, packing rules explain many binary nanoparticle superlattice structures obtained from both spherical and nonspherical particle shapes \cite{bolestalapin,JACSmurray,NLmurray}. Packing rules are also successfully used to design DNA-functionalized gold nanospheres \cite{mirkindnaaniso,mirkindnadesign}. This raises the question: in instances where packing principles can describe observed crystal structures, does that necessarily imply that packing mechanisms are responsible?

For chemically bonded spherical particles, where, e.g. electrostatic forces between oppositely charged colloids or ligand-ligand attraction between functionalized nanoparticles may dominate, packing arguments seem plausible due to the tendency towards close-packed structures. However, when attractive interparticle forces are weak and particles are nonspherical, entropy arising from thermal motion can dominate and invalidate packing rules \cite{Damasceno2012c, dijkstracube,dijkstraentdr,dijkstranonconvex,rossipnas, escobedo_ent,escobedordsq,Escobedo2014, lockkeyent}. Nevertheless, there are examples in both situations where packing rules appear to explain self-assembled structures. Does that imply the crystal \textit{formed} via a packing mechanism? Or is it simply the case that packing rules are useful to rationalize the structure retrospectively, as is the case for molecular packing rules in ordered surfactant systems?

Statistical thermodynamics tells us that free-energy minimization dictates equilibrium structures. In the case of hard particles, free energy minimization is achieved by structures that self-assemble to maximize entropy except in the limit of very high pressures, where they maximize density \cite{VanAnders2014d,VanAnders2014c,Escobedo2014,Frenkel1999,dogic,Manoharan2015b,Frenkel2015}. It is these maximum density (or infinite pressure) structures that are invoked when packings are discussed. It is also this limiting case that offers an explanation of why systems of atoms, molecules, or nanoparticles might order through packing. 

We can answer our questions by comparing for a given system the
self-assembly density, $\eta_\text{assembly}$: the lowest density at which
spontaneous self-assembly is observed, with the ``packing onset density'', $\eta_\text{pack}$: the lowest density at which the system exhibits packing behavior. 
We argue that packing behavior can occur in a finite pressure system if it follows the same asymptotic, infinite pressure behavior as idealized, mathematical packing.
So the question of ``when does matter pack?'' reduces to searching for this asymptotic behavior. We test for the existence of this asymptotic behavior using generalized Maxwell relations derived in the alchemical ensemble first introduced in Ref.~\cite{VanAnders2015}. These generalized Maxwell relations are similar to the usual ones, but defined here in shape space, they can be used to define the packing onset density. One of these generalized relations directly relates the density of ordered structures to the ``alchemical potential'' $\mu$, defined as the change in the alchemical free energy in response to a change in particle shape:
\begin{equation}
 \left(\frac{\partial\mu_i}{\partial P}\right)_{N,T,\alpha_j} =
 \frac{1}{\eta^2}
 \left(\frac{\partial\eta}{\partial\alpha_i}\right)_{N,P,T,\alpha_{j\neq i}}
\label{eq:mxwl}
\end{equation}
Here $\eta$ denotes density, $\alpha$ represents the alchemical (here, shape) variable, $P$ is pressure, $N$ is the total number of particles, and $T$ is temperature responsible for the thermal motion of the particles. The right-hand side of this equation can be computed analytically for systems at the limit of infinite pressure, i.e. densely packed systems, while the left-hand side can be computed via simulations in the isobaric alchemical ensemble \cite{VanAnders2015}. We define $\eta_{\text{pack}}$ as the lowest density that satisfies this generalized Maxwell relation when the right-hand side is calculated at infinite pressure. 
If we find for some system that $\eta_\text{assembly}\approx\eta_\text{pack}$, then this
indicates that the onset of order is consistent with the existence of a global,
dense packing mechanism.
However, if $\eta_\text{assembly}<\eta_\text{pack}$ this
indicates that systems spontaneously order before they pack, and the mechanism
that drives the order is not packing. However, in this case, it is
possible that systems could be quenched to $\eta_\text{pack}$ in a disordered
state, and then subsequently order by packing. To determine
if this is possible, we compare $\eta_\text{pack}$ to random close packing
densities $\eta_\text{rcp}$. If $\eta_\text{pack}>\eta_\text{rcp}$ then we
conclude that a given system cannot be ordered by a packing mechanism.
Above $\eta_{\text{rcp}}$, only entropy maximization can order a system of hard particles. We will show below that in all systems we study $\eta_{\text{pack}} > \eta_{\text{rcp}} > \eta_{\text{assembly}}$, indicating that not only is the spontaneous order not driven by packing, but also the systems cannot be ordered by packing. 

We also pose the following, related question: When can packing arguments be used for the inverse problem of predicting the thermodynamically optimal particle shape for a particular structure? In other words, when---if ever---is the space-filling hard shape of a target crystal structure thermodynamically optimal for self-assembling that crystal?

\begin{figure}
\includegraphics[width=0.45\textwidth]{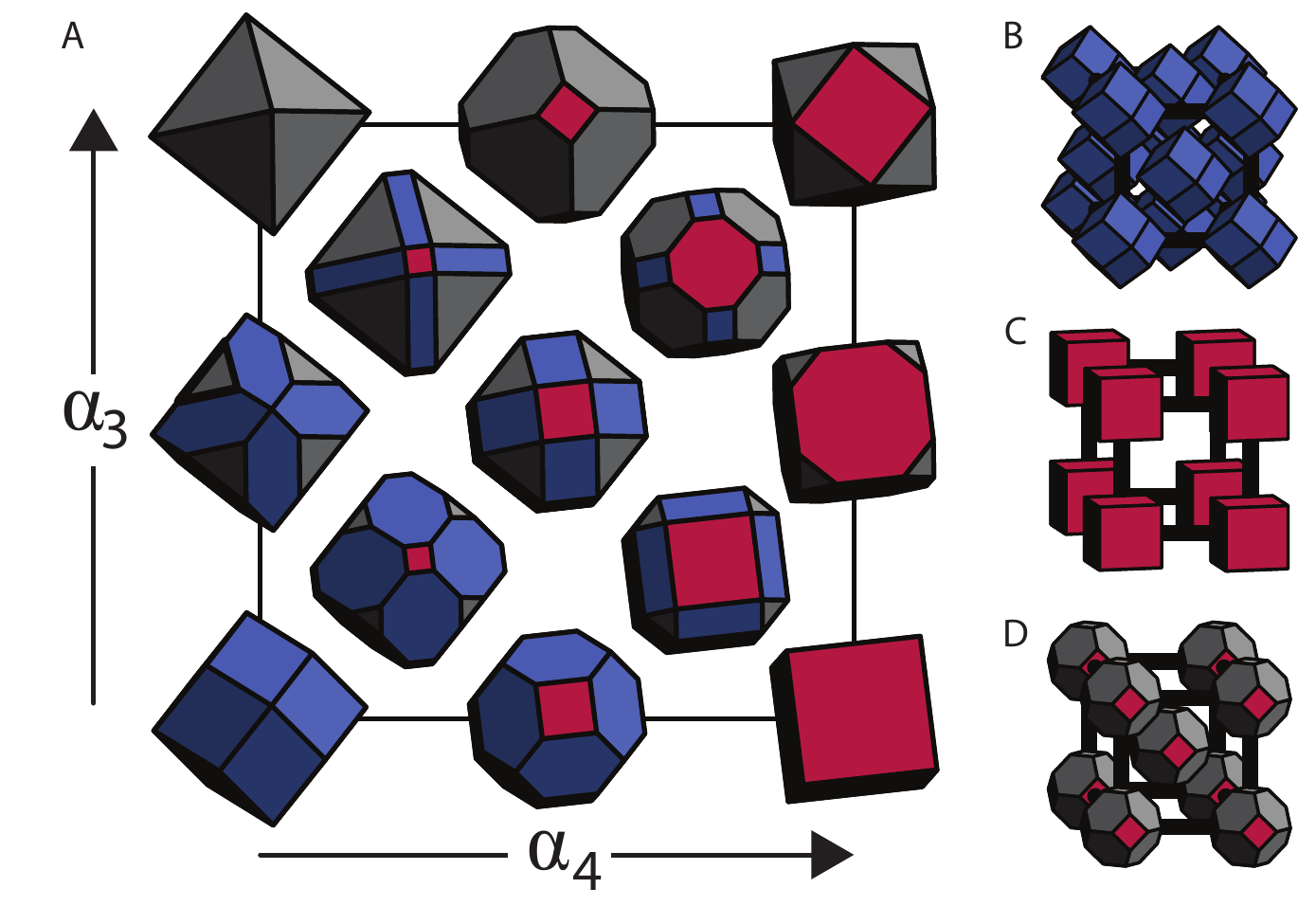}
\caption{\textbf{Shape Family and Structures}. We rely on a previously defined parameterization \textbf{(A)} which continuously maps two values, $\alpha_3$ and $\alpha_4$, to convex polyhedra. This parameterization, here known as the $\Delta_{423}$ family, contains the space-filling shapes for \textbf{(B)} FCC, \textbf{(C)} SC and \textbf{(D)} BCC. 
}
\label{fig:Delta423}
\end{figure}

\section*{Approach}
To understand whether packing is driving self-assembly or fundamental to particle design, we study the most likely systems for this to be the case: idealized, perfectly hard, convex shapes. We consider three common structures: face-centered cubic (FCC), simple cubic (SC), and body-centered cubic (BCC) and their corresponding space-filling (Voronoi) shapes: rhombic dodecahedron, cube,  and truncated octahedron. It is well known that, for each of these shapes, the corresponding structure is the only thermodynamic equilibrium assembly and the densest packing (at $\eta = 1$ by definition) \cite{Damasceno2012a,dijkstrasuperballs,dijkstracube,Damasceno2012c,dijkstratcube,VanAnders2014c}. Yet, we show that, in each case, the onset of packing behavior occurs at a higher density than $\eta_{\textrm{\text{rcp}}}$, and therefore the observed structures cannot be ordering via a packing mechanism. Moreover, we show that the space-filling shape for all three structures is never thermodynamically optimal except at $\eta = 1$. From these two findings, we argue that there is only a correlation, not a causal relationship, between the observed thermodynamically assembled structures and those rationalized by packing arguments.

Hard Particle Monte Carlo (HPMC) simulations \cite{Anderson2016} of the self assembly of FCC, SC, and BCC crystals were carried out for a family of spheric triangle group invariant particle shapes, $\Delta_{423}$ (Fig.~\textbf{\ref{fig:Delta423}A}), which includes each crystal's space-filling particle, as shown in Fig.~\textbf{\ref{fig:Delta423}(B--D)}, but also, importantly, sets of truncated versions of those shapes that are nearby in shape space. This shape family maps two values, $\alpha_3$ and $\alpha_4$ to convex polyhedra, with $\alpha_3, \alpha_4 \in [0,1]$. The space-filling shapes for FCC, SC, and BCC are defined at $\left(\alpha_4, \alpha_3\right) = (0,0)$ (rhombic dodecahedron), $\left(\alpha_4, \alpha_3\right) = (1,0)$ (cube), and $\left(\alpha_4, \alpha_3\right) = (\frac{2}{3},1)$ (Archimedean truncated octahedron), respectively. 

\begin{figure}
\includegraphics[width=0.45\textwidth]{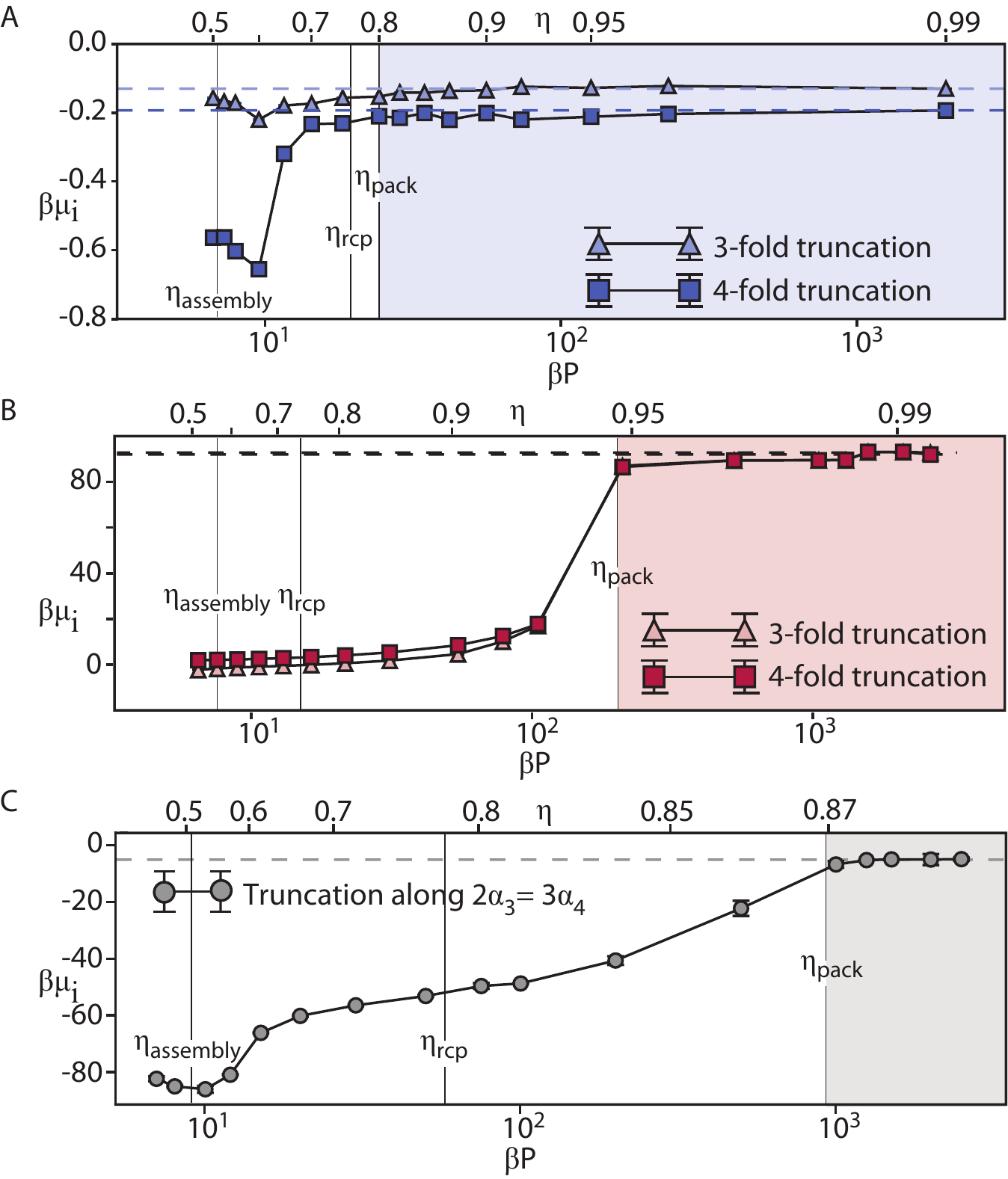}
\caption{\textbf{Onset of Asymptotic Packing Behavior in \textbf{(A)} FCC, \textbf{(B)} SC, and \textbf{(C)} BCC.} We utilize the generalized Maxwell relation in \eqref{eq:mxwl} to extract the onset of asymptotic packing behavior for \textbf{(A)} FCC, \textbf{(B)} SC, and \textbf{(C)} BCC with respect to their space-filing particles. The shaded region in each figure denotes the density-pressure regime where the system is found to be ``packing'', i.e. where structure formation is driven by packing principles. By comparison, $\eta_{\text{pack}}$ is much higher than either $\eta_{\text{assembly}}$ and $\eta_{\text{rcp}}$ shown for all three systems. For space-filling particles within FCC, SC, and BCC, $\eta_{\text{assembly}}\approx$ 0.5--0.55 \cite{escobedo,dijkstracube}, and $\eta_{\text{rcp}}\approx$ 0.76, 0.74, and 0.78, respectively, calculated using methods from Refs.~\cite{kamienliurcp, songmasoneqs}.}
\label{fig:AP}
\end{figure}

To compute the packing onset density, we utilized analytical constructions of putative densest packings reported in Ref.~\cite{Chen2014} for the entire $\Delta_{423}$ shape family, giving $\eta$ as a function of $\alpha$, to evaluate the right-hand side of \eqref{eq:mxwl} in the infinite pressure limit. We evaluated the left-hand side of \eqref{eq:mxwl} at finite pressure using NPT$\alpha$ HPMC simulations at varying pressures. We also performed simulations in the NVT$\mu$ ensemble to find the thermodynamically optimal shape for FCC, SC, and BCC as a function of density. Additional details and derivations for the parameterization of the $\Delta_{423}$ shape family, extended ensembles, and the simulations conducted can be found in the Methods Section and in Supplementary Information.

\begin{figure*}
\includegraphics[width=0.95\textwidth]{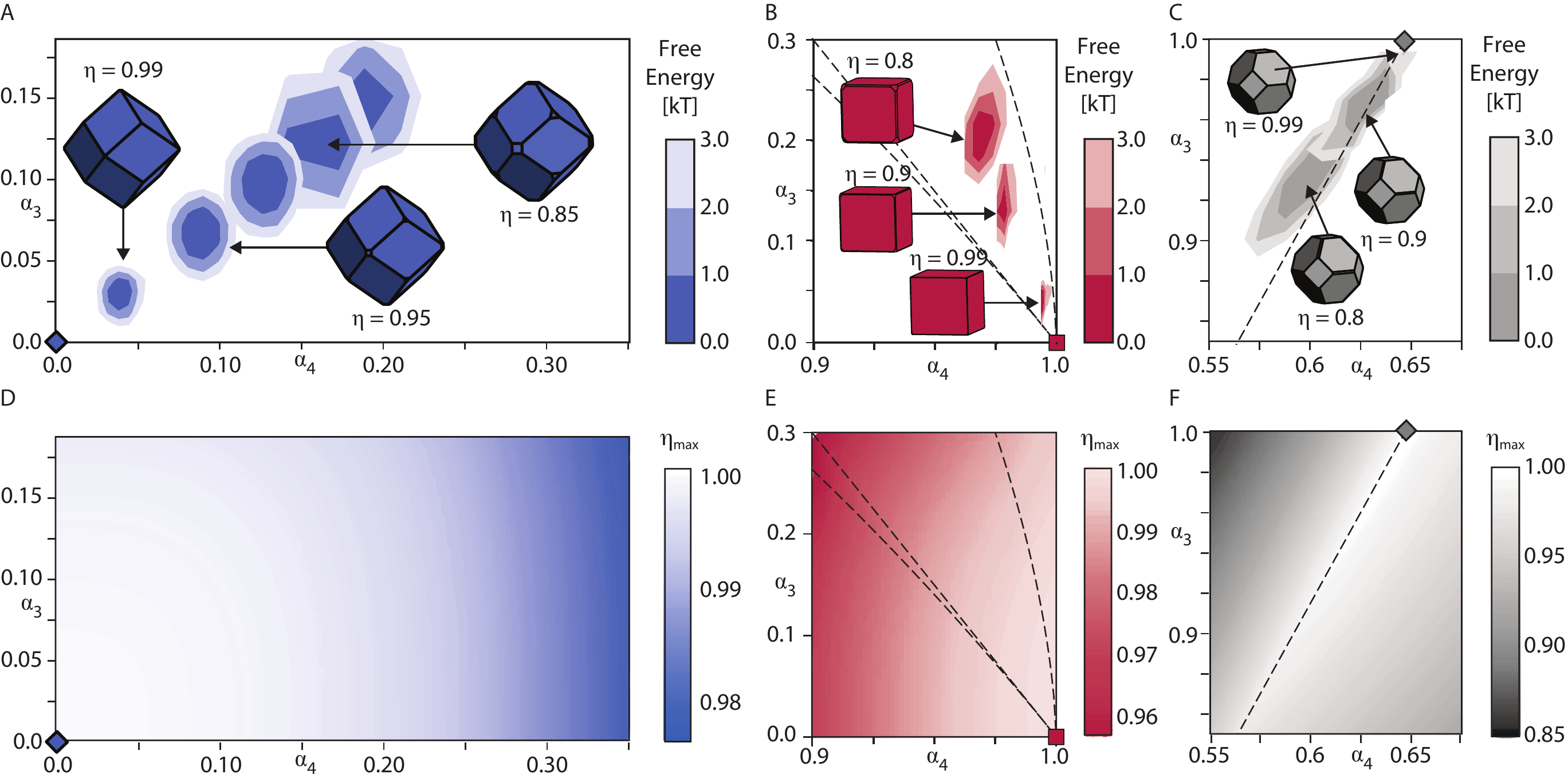}
\caption{\textbf{Optimal Particle Shape for \textbf{(A)} FCC, \textbf{(B)} SC, and \textbf{(C)} BCC} The perfect space-filling (Voronoi) shape, denoted by a solid polygon in each figure, is different from any of the optimal shapes determined from simulation, even up to densities of $0.99$. \textbf{(D--F)} The densest packing surface reported in Ref.~\cite{Chen2014} in the regions closest to the space-filling shapes for \textbf{(D)} FCC, \textbf{(E)} SC, and \textbf{(F)} BCC The dotted lines in \textbf{(B--C), (E--F)} denote discontinuities in the derivative of the dense filling surface.}
\label{fig:OS}
\end{figure*}

\section*{Results and Discussion}
The computed alchemical potential as a function of pressure is plotted in Fig.~\textbf{\ref{fig:AP}} for \textbf{(A)} FCC, \textbf{(B)} SC, and \textbf{(C)} BCC. Particle shape is fixed to that of the space-filling particle in each case.
Asymptotic behavior extracted from analytical results reported in Ref.\ \cite{Chen2014} reveals that in all three cases, asymptotes have zero slope in the limit of infinite pressure, indicated with horizontal dotted lines in each panel. We distinguish the onset of packing behavior with a vertical line in each panel, at $\eta_{\text{pack}} = $ 0.80, 0.95, and 0.87 for FCC, SC, and BCC, respectively, as the lowest densities for which the left-hand side of \eqref{eq:mxwl} approaches the infinite pressure limit of the right-hand side. We also indicate both the assembly and random close packing densities for FCC, SC, and BCC: $\eta_{\text{assembly}}\approx$ 0.5--0.55 \cite{escobedo,dijkstracube}, and $\eta_{\text{rcp}} = $ $0.76\pm0.03$, $0.74\pm0.03$, $0.78\pm0.03$, respectively. The values of $\eta_{\text{rcp}}$ are approximate upper limits, computed from methods described in Refs.~\cite{kamienliurcp, songmasoneqs}. In every system, $\eta_{\text{pack}} > \eta_{\text{rcp}} > \eta_{\text{assembly}}$. Our results indicate that none of the systems investigated here order via a packing mechanism; rather, they indicate only that systems can self-assemble into the same structures that correspond to packings. 

Moreover, results in Fig.~\textbf{\ref{fig:OS}(A--C)} indicate that packing cannot predict ideal particle shapes for self-assembly because the perfect space-filling shape
is never thermodynamically preferred away from $\eta = 1$. Even at $\eta = 0.99$, there is a tiny difference in shape that, in each case, produces a non-trivial difference in free energy (up to $>3kT$) between identical crystals comprised of the thermodynamically optimal shape and those comprised of the space-filling shape. However, we do find that features in the global dense packing landscape do \textit{generally} correlate with features found in the optimal particle shapes, as seen by comparing the optimal shape (Fig.~\textbf{\ref{fig:OS}(A--C)}) with the corresponding densest packing landscape from Ref.~\cite{Chen2014}~(recreated in Fig.~\textbf{\ref{fig:OS}(D--F)}). Thus, while densest packing arguments do not predict the optimal shape for self-assembly, the densest packing landscape may provide guidance in determining optimal particle shape.

\section*{Conclusions} 
Although packing arguments are often used successfully in nanoparticle and colloidal assembly, they often fail to explain experimental and computational observations. Our findings demonstrate that the use of packing arguments to rationalize observed structures or design particles to achieve target structures may not be well founded, even when the observed structure is the same as one would get from packing. Because one would expect packing principles---if they do hold---to hold for hard particles, our finding raises the question: is the apparent success of Pauling's packing principles for atomic systems also a spurious correlation? It could also be that the imperfect hardness of atoms and molecules makes them more amenable to dense packing as a mechanism. This counterintuitive possibility would beg for further understanding as the initial reasons for applying packing arguments were based on the existence of steep sterically repulsive interaction potentials that are nearly hard. Combining the approach for studying sphere packings developed in Ref.~\cite{kallusulam} with the generalized Maxwell relation \eqref{eq:mxwl} could give insight into that question. 
In providing new, thermodynamic formulations with which to investigate the packing of hard shapes, we offer alternative approaches to the ubiquitous but notoriously difficult set of general packing problems \cite{bolestalapin, Lagarias2012}. 

Our findings also show that, even near densities of 1, entropic contributions cannot be ignored. Small, stabilizing entropic contributions one might have guessed to be irrelevant can arise from nearly infinitesimal shape modifications, such as small truncations of vertices or edges of polyhedral nanoparticles. This means the space-filling shape is never thermodynamically optimal for self-assembling its corresponding structure, at least for hard colloidal particles. Our observations suggest that heroic efforts to synthesize perfectly shaped, space-filling particles to achieve the corresponding target structure are unnecessary, and that the entropy gained from slight imperfections may actually facilitate assembly. 

\section*{Methods}

Here we present a brief overview of the methods used to generate the data given in the main text. Further details and derivations can be found in the Supplementary Information.

\subsection{Shape Parameterization}
Spheric triangle-group families are generated by the intersection of sets of symmetric planes \cite{Chen2014}.
They are parameterized by $\alpha_i$ parameters between 0 and 1, where $\alpha_i$ encodes the
inverse distance of the i\textsuperscript{th}-fold symmetric planes from the particle center. Any $\alpha_i$ parameter can be more readily understood
as the truncation of an i\textsuperscript{th} fold axis of symmetry for the particle shape.

The shape family studied here is generated by intersections of planes perpendicular to the directions of the 4-fold, 2-fold, and 3-fold axes of rotational symmetry for a rhombic
dodecahedron, shown in Fig.~\textbf{\ref{fig:Delta423}A} as red, blue, and grey, respectively, and is therefore named the 423 family of polyhedra ($\Delta_{423}$).
$\Delta_{423}$ is parameterized by three values, $\alpha_2$, $\alpha_3$ and $\alpha_4$, in the manner described above. Planes perpendicular to the 2-fold direction remain fixed, as does $\alpha_2$, so discussion is restricted to $\alpha_3$ and $\alpha_4$. 
We restrict our exploration to a shape space with restricted point group symmetry due to (i) geometric reasoning about shape features that
lead to optimal thermodynamic behavior (ii) crystal growth processes that
determine particle symmetries in nanoscale and synthesis protocols (e.g.
 \cite{Sun13122002,saurogach,skrabalak,skrabalak}).

\subsection{Simulation Methods}
We simulate our shapes in the \textit{alchemical ensemble} using the 
digital alchemy (DA) framework \cite{VanAnders2015}. DA is a statistical mechanics simulation technique that employs
thermodynamic ensembles extended into alchemical (here, particle shape) space by one or more
dimensions, allowing fluctuations in the alchemical space or corresponding conjugate
alchemical potential(s). This extended (``alchemical'')
ensemble has the partition function \cite{VanAnders2015}
\begin{equation}
 \Zcal = e^{-\beta F} = \sum_\sigma
 e^{-\beta(H-\sum_i \mu_i N \alpha_i-k\Lambda)} \; ,
 \label{eq:zcal}
\end{equation}
where $\sigma$ labels microstates, $H$ is the Hamiltonian, $\alpha_i$ are the
alchemical parameters describing model-specific particle shape attributes (detailed above),
$\mu_i$ are thermodynamically conjugate alchemical potentials, $N$ is the number
of particles, $\Lambda$ is a structural design criterion that initially keeps the system in an FCC, SC, or BCC crystal structure,
and $k$ is the strength of the coupling to $\Lambda$. 

We employed DA through the simulation method alchemical Monte Carlo (Alch--HPMC) \cite{VanAnders2015}.
In Alch--HPMC simulations, $\mu_i$ is held constant and particle position and orientation moves are accepted with standard acceptance criteria \cite{Frenkel2002}. Unbiased
shape moves (hereafter $\mu_i=0$) are performed such that all particles in the system
change simultaneously from a shape described by alchemical parameter $\alpha$ to a shape
described by $\alpha'$, with probability
\begin{equation}
 \pi = \min \left( 1,
 \frac{(\det(I_{\alpha}))^{N/2}}{(\det(I_{\alpha'}))^{N/2}}
 e^{-\beta(U_{\alpha}-U_{\alpha'})}\right)
 \label{eq:metropolis}
\end{equation}
where $U$ is the potential energy and $I$ is the anisotropic particle moment of
inertia tensor. See Supplementary Information and \cite{VanAnders2015} for details. For the hard particle systems studied here the potential energy
(and $\Delta U$) vanishes for all valid, non-overlapping particle configurations
so that $\pi = 0$ for a microstate in which any particles overlap.

We used DA and Alch--HPMC in two ways: 
(i) we computed the expectation value $\langle\alpha_i\rangle$ as a function of packing fraction in the $NVT\mu_i$ ensemble, and (ii) we performed Alch--HPMC within the $NPT\alpha_i$ ensemble, obtained by Legendre transforming the $NVT\mu_i$ ensemble twice, to calculate the alchemical potential $\mu_i$.
All simulations were run with systems of 500 or more particles. See Supplementary Information for numerical details and state points for $NVT\mu_i$ and $NPT\alpha_i$ simulations.

In (i), we initialized independent simulations with distinct shapes, taking
$\Lambda$ to be the potential energy function of an Einstein crystal for the
target structure (FCC, SC or BCC at some density). We maintained non-zero $k$ during initialization only to
ensure the system did not transition out of the target structure. All data
were collected on equilibrated systems with $k=0$; results were validated by
directly computing the free energy \cite{Frenkel1984} for selected state points
in $NVT\mu_i$ simulations. 

In (ii), simulations were used to evaluate alchemical potentials for space-filling shapes using the thermodynamic relation
\begin{equation}
 \mu_i = -\frac{1}{N}\left(\frac{\partial F}{\partial
 \alpha_i}\right)_{N,\eta,T,\mu_{j\neq{i}}}
\label{eq:alchpo}
\end{equation}
Our Alch--HPMC algorithm recorded the acceptance ratio for small trial moves from the space-filling shape, without performing such moves. 
We evaluated \eqref{eq:alchpo} numerically using the Bennett acceptance ratio method \cite{bennettacpr}, which is described as it applies to the alchemical potential in Ref.~\cite{VanAnders2015} and employs a finite differencing method published in Ref.~\cite{fornberg}.

In this ensemble, we derive a Maxwell relation between alchemical potential $\mu$ and packing fraction $\eta$:
\begin{equation}
 \left(\frac{\partial\mu_i}{\partial P}\right)_{N,T,\alpha_j} =
 \frac{1}{\eta^2}
 \left(\frac{\partial\eta}{\partial\alpha_i}\right)_{N,P,T,\alpha_{j\neq i}}
\end{equation}
We used this expression to relate the high pressure asymptotic behavior of the alchemical potential to dense packing surfaces that have been analytically computed in the literature \cite{Chen2014}.

Specifically, we consider systems to exhibit packing
behavior when the slope of the alchemical potential approaches the infinite pressure asymptotic limit of the dense packing surface given by \eqref{eq:mxwl}. Relevant data \cite{Chen2014} has been recreated in the Supplementary Information according to our variable notation.

All simulations were performed with a hard particle Monte Carlo (HPMC) \cite{Anderson2016} extension to HOOMD-Blue \cite{Anderson2008,Anderson2013e}, which we further extended to allow Alch--HPMC moves. Runs were partially performed on XSEDE computing resources \cite{xsede}. The data management for this publication was supported by the signac data management framework \cite{Adorf2016, signacSoftware}. Details on statistical analyses can be found in Supplementary Information.

\section*{Acknowledgements}
This work used the Extreme Science and Engineering Discovery Environment (XSEDE), which is supported by National Science Foundation grant number ACI-1053575; XSEDE award DMR 140129.
R.K.C. acknowledges support from the University of Michigan Rackham Merit Fellowship program and National Science Foundation, Division of Materials Research Award No. DMR 1120923.
P.M.D. would like to acknowledge support by the National Science Foundation, Emerging Frontiers in Research and Innovation Award No. EFRI-1240264.
This material is based upon work supported by (in part by) the U.S. Army Research Office under Grant Award No. W911NF-10-1-0518
This work was partially supported by a Simons Investigator award from the Simons Foundation to S.C.G.

We thank J. Dshemuchadse, E. Harper, Y. Geng, and C.X. Du for helpful conversations.

\end{document}